\documentclass[aps, prl,twocolumn]{revtex4}

\usepackage{color}

\usepackage{graphicx}

\usepackage{amsmath}

\usepackage{amssymb}

\usepackage{soul}

\usepackage{float}

\usepackage{dcolumn}

\usepackage{bm}

\usepackage{epstopdf}

\usepackage{color}

\newcommand{\FC}{f_\mathrm{C}}

\newcommand{\FA}{f_\mathrm{A}}

\newcommand{\FSC}{f_\mathrm{SC}}

\newcommand{\FSA}{f_\mathrm{SA}}

\newcommand{\THi}{\theta_\mathrm{i}}

\newcommand{\Ei}{E_\mathrm{i}}

\begin{document}

\title{Graphene Reflectarray Metasurface for Terahertz Beam Steering and Phase Modulation}

\author{M. Tamagnone $^{1,2,*,\dagger}$, S. Capdevila$^{1,\dagger}$, A. Lombardo$^{3}$, J. Wu$^{3}$, A. Centeno$^{4}$, A. Zurutuza$^{4}$, A. M. Ionescu$^{5}$, A. C. Ferrari$^{3}$, J. R. Mosig$^{1}$}

\affiliation{$^{1}$Laboratory of Electromagnetics and Antennas, \'Ecole Polytechnique F\'ed\'erale de Lausanne, Lausanne, Switzerland}

\affiliation{$^{2}$Harvard John A. Paulson School of Engineering and Applied Sciences, Harvard University, Cambridge, Massachusetts 02138, USA}

\affiliation{$^3$Cambridge Graphene Centre, University of Cambridge, 9 J.J. Thompson Avenue, Cambridge CB3 OFA, UK}

\affiliation{$^4$Graphenea SA, 20018 Donostia-San Sebasti\'an, Spain}

\affiliation{$^5$Nanoelectronic Devices Laboratory, \'Ecole Polytechnique F\'ed\'erale de Lausanne, Lausanne, Switzerland}

\affiliation{$^*$Corresponding authors}

\affiliation{$^{\dagger}$These Authors contributed equally}

\begin{abstract}

We report a THz reflectarray metasurface which uses graphene as active element to achieve beam steering, shaping and broadband phase modulation. This is based on the creation of a voltage controlled reconfigurable phase hologram, which can impart different reflection angles and phases to an incident beam, thus replacing bulky and fragile rotating mirrors used for terahertz imaging. This can also find applications in other regions of the electromagnetic spectrum, paving the way to versatile optical devices including light radars, adaptive optics, electro-optical modulators and screens.

\end{abstract}

\maketitle

\begin{figure*}

\centerline{\includegraphics[width=180mm]{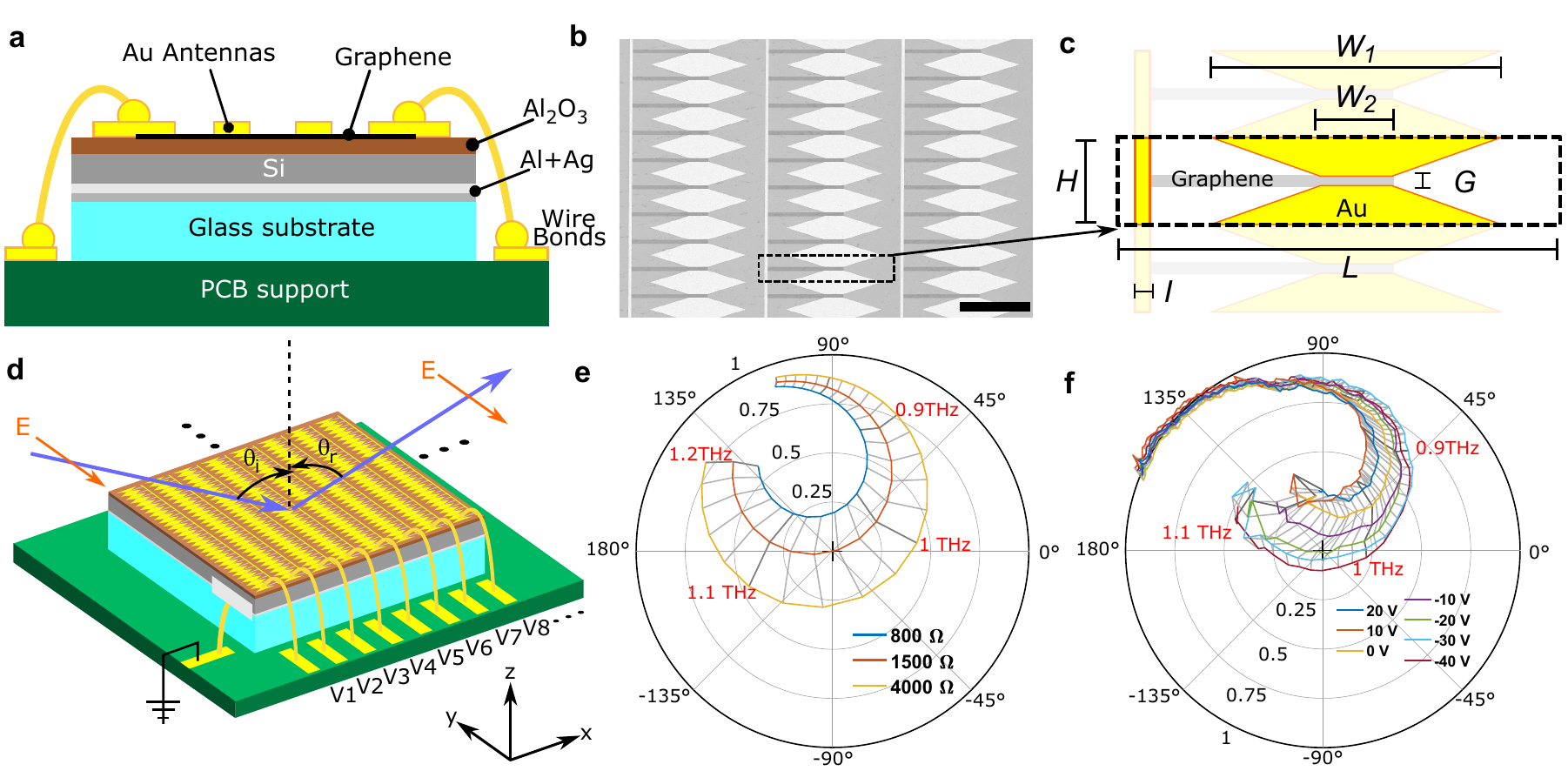}}

\caption{\label{fig:Figure1} \textbf{a}, Cross section of the device mounted on a printed circuit board (PCB) and wire-bonded. Thickness of the layers from bottom to top: glass substrate 525 $\mu$m, evaporated Al 100 nm; Ag 100 nm, Si 20 $\mu$m, Al$_2$O$_3$ 200 nm, Au 100 nm; 5 nm Cr adhesion layer). \textbf{b}, Scanning electron microscope (SEM) picture of a representative device (scale bar 50 $\mu$m). \textbf{c} Unit cell. width L = 100 $\mu$m, height H = 20 $\mu$m, antenna gap G = 3 $\mu$m, first trapezoid base W$_1$ = 70 $\mu$m, second trapezoid base W$_2$ = 7 $\mu$m, width of interconnecting line I = 2$\mu$m. The final size of the array area is 8$\times$8 mm$^2$, hence 80x400 cells. \textbf{d}, characterization of the electric modulation of the complex reflection coefficient. Light is s-polarized and incident with an angle $\THi=45^\circ$. In this case, all columns are driven with the same voltage ($V_1 = V_2 = V_3 = ... = V_N$), therefore SLG has the same $\sigma$ in all the cells. Since the cells are smaller than half wavelength, light is reflected specularly ($\theta_r=45^\circ$) and the reflection coefficient depends on the frequency and $\sigma$.\textbf{e}, simulations and \textbf{f}, measurements of the complex reflection coefficient of the array. The colored lines in the polar plots represent the evolution of the reflective index with frequency (indicated in red), and each curve represents a different SLG impedance or biasing voltage. The transverse grey lines join data points measured at the same frequency.}

\end{figure*}

\begin{figure*}

\centerline{\includegraphics[width=180mm]{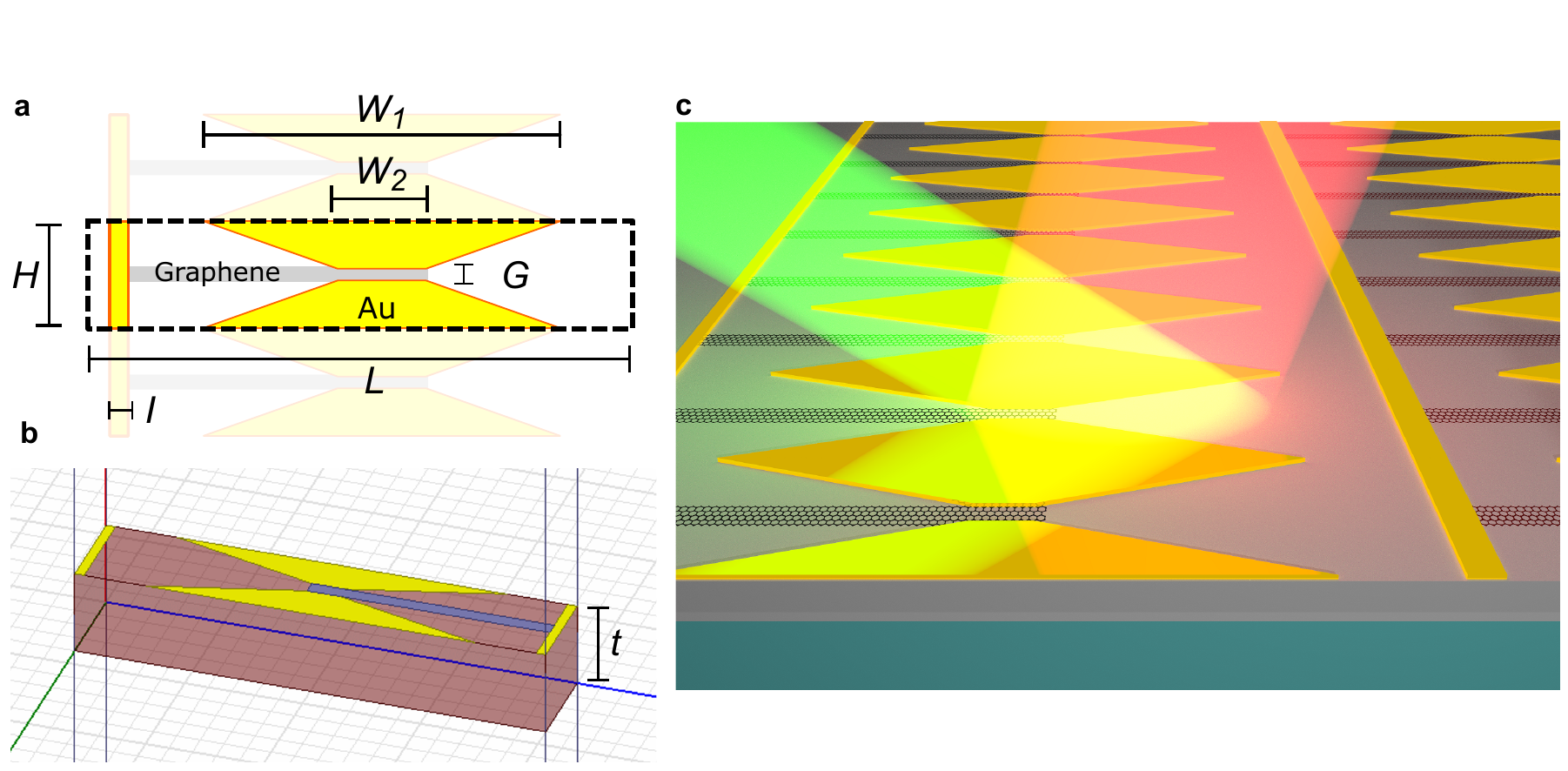}}

\caption{\label{fig:SupplFigure4}\textbf{a}, Unit cell dimensions after optimization. L = 100 $\mu$m, H = 20 $\mu$m, G = 3 $\mu$m, W$_1$ = 70 $\mu$m, W$_2$ = 7 $\mu$m, I = 2 $\mu$m. \textbf{b}, Screenshot of the unit cell simulation setup (the used software is Ansys HFSS). \textbf{c}, Artistic view of the final array.}

\end{figure*}

\begin{figure*}

\centerline{\includegraphics[width=180mm]{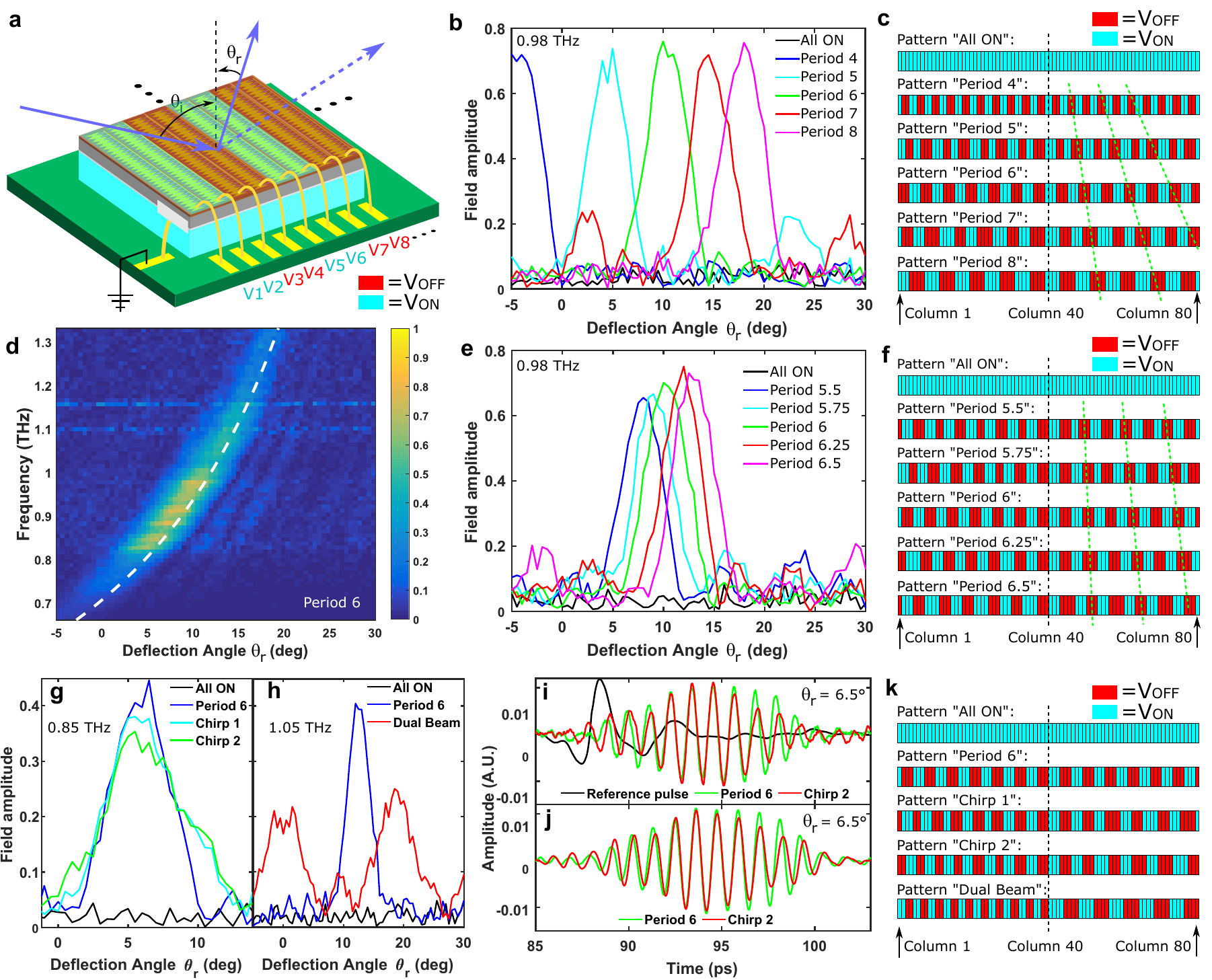}}

\caption{\label{fig:Figure2} \textbf{a}, Beam steering principle. Columns are alternatively set ON and OFF ($V_\mathrm{ON}$ = 26 V, $V_\mathrm{OFF}$ = -44 V). \textbf{b-c} coarse beam steering obtained using periodic patterns.\textbf{b} Beam profiles at 0.977 THz, where the deflected field is higher. Amplitudes are normalized to the largest deflected field for any angle and frequency. \textbf{c}, control voltage patterns used in \textbf{b}. \textbf{d} angular dispersion of the beam for different frequencies. The dashed white line is the expected theoretical dispersion from Eq.\ref{eqn:steer}. \textbf{e-f}, fine beam steering from quasi-periodic patterns. \textbf{e} beam profile at 0.977 THz. \textbf{f} control voltage patterns used in \textbf{e}. \textbf{g} beam broadening using chirped patterns. \textbf{h} dual beam operation. \textbf{i} time response of steered beam, compared to the pulse (not in scale) of our THz time domain system (TDS). \textbf{j} time response of steered beam de-convoluted to remove the measurement pulse shape. Both \textbf{i} and \textbf{j} show a standard deflected beam (P = 6) and one obtained with a chirped pattern. The time response of the chirped pattern is also chirped: the red and green traces are aligned in the centre, but the chirped pattern is delayed with respect to the non-chirped one both at the end and at the beginning of the pulse. \textbf{k} chirped and dual beam control patterns.}

\end{figure*}

\begin{figure*}

\centerline{\includegraphics[width=100mm]{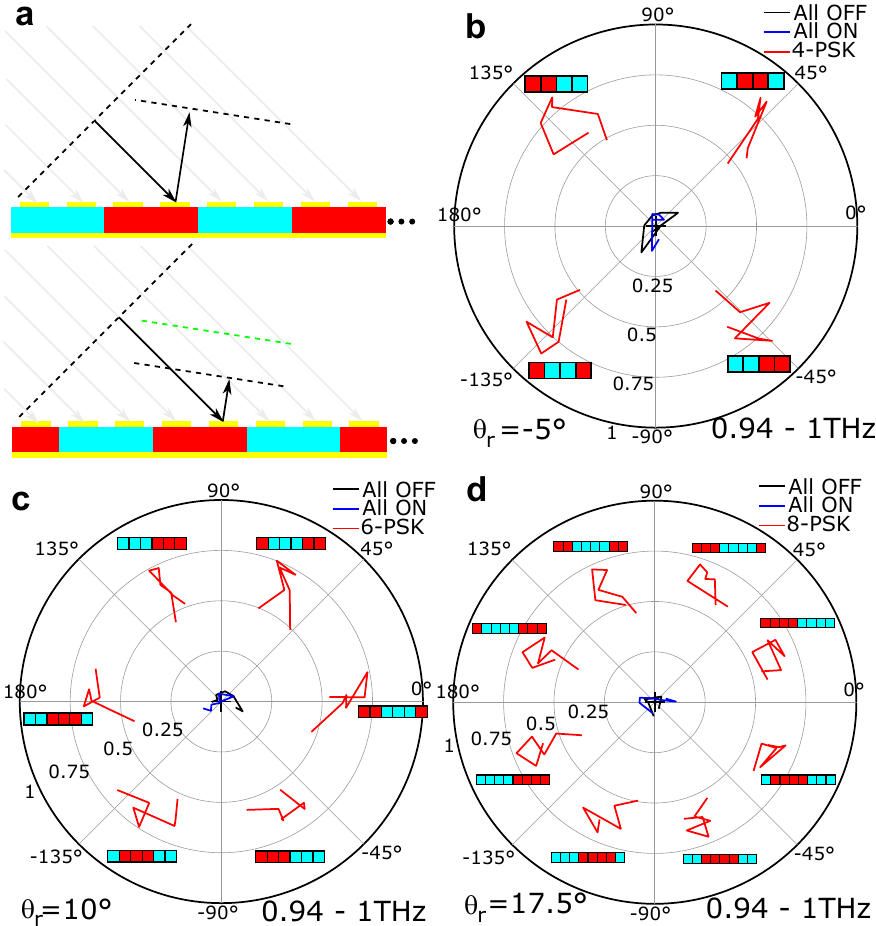}}

\caption{\label{fig:Figure3} \textbf{a} Geometric discrete phase modulation principle. The solid black line indicates how the delay of the deflected beam can be modified by shifting the control voltage pattern. \textbf{b-d} measured geometrical phase modulation by shifting a periodic pattern with P=4, 6, 8. The resulting phase shift keying (PSK) constellations on the complex plane are shown on a polar plot with lines that represent the symbol evolution from 0.94 to 1THz. Details on background removal are given in Method \ref{met:measurement}. Close to each symbol is the supercell used to generate it. Each measurement is done at the angle of maximum intensity, see Fig.\ref{fig:Figure2}b.}

\end{figure*}

Metasurfaces are planar devices based on a periodic or quasi-periodic bi-dimensional array of cells (typically dielectric or metallic elements placed on a layered substrate) capable of manipulating impinging light to obtain various functionalities, such as focusing\cite{khorasaninejad2016metalenses,headland2016dielectric}, beam steering and shaping\cite{HumPerruisseau-Carrier2014,DebogovicBartolicPerruisseau-Carrier2014,CarrascoTamagnoneMosigEtAl2015}, unilateral propagation\cite{TamagnoneFallahiMosigEtAl2014}, polarization control\cite{mueller2017metasurface,NiuWithayachumnankulUpadhyayEtAl2014}, frequency filtering\cite{HasaniTamagnoneCascanteEtAl2016,mittra1988techniques}, non-linear phenomena\cite{lee2014giant}, and dynamic modulation\cite{sun2016optical,TamagnoneFallahiMosigEtAl2014,LiYu2013,Sensale-RodriguezYanRafiqueEtAl2012,kim2016vanadium,sherrott2017experimental,miao2015widely,YaoShankarKatsEtAl2014}. When optically tuneable materials are embedded in the cells, metasurfaces can be designed to dynamically steer a beam in different directions. This can be achieved at microwave frequencies in reflectarray (RA) metasurfaces using, e.g., micro-electrical-mechanical-systems (MEMS)\cite{HumPerruisseau-Carrier2014,DebogovicBartolicPerruisseau-Carrier2014}, or voltage controlled capacitors\cite{HumPerruisseau-Carrier2014,RodrigoJofrePerruisseau-Carrier2013}. Extending beam steering to THz, infrared and visible frequencies is challenging due to the scarcity of compact tuneable elements operating at shorter wavelengths.

This technological issue can be solved by using single layer graphene (SLG), which is an ideal material for photonics and optoelectronics due to its rich physics and gate-tuneable optical properties\cite{BonaccorsoSunHasanEtAl2010}. Compared to bulk materials, the possibility of inducing high carrier densities in SLG is the key to achieve optical tuneability both for optical intraband processes \cite{Sensale-RodriguezYanRafiqueEtAl2012,TamagnoneMoldovanPoumirolEtAl2016b} and for interband processes\cite{LiuYinUlin-AvilaEtAl2011} (which are relevant for optoelectronic modulators and photodetectors\cite{KoppensMuellerAvourisEtAl2014}). Furthermore, high mobility of SLG allows mid infrared plasmon-polaritons, also tuneable by gating\cite{rodrigo2015mid,YanLowZhuEtAl2013}.

SLG is ideally suited to modulate terahertz waves because of its high mobility and easy integration on Si technology. The mobility is linked to the massless nature of carriers in SLG and allows for a larger conductivity tuneability range than Si, for a given carrier density interval \cite{Sensale-RodriguezYanRafiqueEtAl2012}. Unlike radio-frequency MEMS\cite{HumPerruisseau-Carrier2014,DebogovicBartolicPerruisseau-Carrier2014}, SLG does not require packaging \cite{Sagade2015}, its switching speed is several order of magnitude faster\cite{LiuYinUlin-AvilaEtAl2011} and it does not suffer from reliability issues\cite{HumPerruisseau-Carrier2014,DebogovicBartolicPerruisseau-Carrier2014}. Tuneable capacitors, instead, are dominated by resistive losses above the microwave range\cite{HumPerruisseau-Carrier2014}. Thus, SLG is an ideal choice for THz modulation\cite{sherrott2017experimental,miao2015widely,Sensale-RodriguezYanRafiqueEtAl2012}.

We report a THz reflectarray metasurface exploiting SLG as active element to achieve beam steering, shaping and broadband phase modulation. Our device achieves dynamical beam steering thanks to an array of cells (including metal and gated SLG) built on a reflective substrate, see Fig.\ref{fig:Figure1}a. This consists of a dielectric spacer layer (20 $\mu$m float zone Si, with dielectric constant 11.7 at 1 THz\cite{headland2016dielectric}), on a metallic reflective film (140 nm Ag on 60 nm Al)\cite{TamagnoneMoldovanPoumirolEtAl2016b,HasaniTamagnoneCascanteEtAl2016}. An additional AlO$_2$ layer is used to gate the SLG, allowing for the dynamical tuning of its conductivity and hence of its optical behavior at THz frequencies. High resistivity Si (for our sample the resistivity is 10 k$\Omega\cdot$cm) is transparent at THz frequencies\cite{HasaniTamagnoneCascanteEtAl2016}, but allows injected carriers to charge the gate capacitance and tune SLG via field effect\cite{Sensale-RodriguezYanRafiqueEtAl2012}. The unit cell in Figs.\ref{fig:Figure1}b,c is inspired by bow tie antennas\cite{Balanis2005}, with two trapezoidal Au elements that concentrate the impinging electromagnetic energy in a 3 $\mu$m narrow gap where SLG is placed.

Fig.\ref{fig:SupplFigure1} in Methods M1 illustrates the RA substrate fabrication process flow. Our device requires a substrate comprising a reflective conductive ground plane and a dielectric spacer with thickness in the order of a quarter wavelength (in the material itself). We achieve this by using high resistivity Si as the dielectric spacer. An anodic bonding process is used to bond a metallic reflective layer (Ag + Al) to a supporting glass substrate. Fig.\ref{fig:SupplFigure2} in Methods M1 summarizes the fabrication of the RA starting from the substrate chips.

The beam steering RA can work as intended only if the cell can modulate its reflection coefficient between two values ($\Gamma_\mathrm{ON}$, $\Gamma_\mathrm{OFF}$) with a phase modulation $\pi$ (see Methods \ref{met:radpattheo}). The amplitude of the reflection coefficient should be maximized and remain constant in the two states. This is equivalent to creating a metasurface where each cell has a tuneable surface impedance, since the reflection coefficient $\Gamma$ and the surface impedance $Z_S$ are related by\cite{Balanis2005,TamagnoneMoldovanPoumirolEtAl2016b}:

\begin{equation}\label{seqn:surfaceimped}
\Gamma=\frac{Z_S-\eta}{Z_S+\eta}
\end{equation}

where $\eta = \sqrt{\mu_0\varepsilon_0^{-1}} \simeq 377\:\Omega$ is the free space impedance, $\mu_0$ is the vacuum magnetic permeability and $\varepsilon_0$ is the vacuum dielectric permittivity. The cell can then be designed to obtain suitable values of $Z_\mathrm{S}$ starting from the SLG's sheet impedance $Z_\mathrm{g} = \sigma^{-1}$, where $\sigma$ is SLG's conductivity. This can be changed via electric field gating between a maximum ($\sigma_\mathrm{ON}$) and minimum ($\sigma_\mathrm{OFF}$). It is possible to control the reflection phase in a binary way (two opposite values of the phase) if the metasurface is designed to have complete absorption ($\Gamma=0$) for $Z_\mathrm{S}=\sqrt{Z_\mathrm{S,ON}\cdot Z_\mathrm{S,OFF}}$. Because $\Gamma=0$ implies $Z_\mathrm{S}=\eta$, the approximate design condition becomes $\sqrt{Z_\mathrm{S,ON}\cdot Z_\mathrm{S,OFF}}=\eta$. The geometric average is used here, to ensure that $\Gamma_\mathrm{ON}=-\Gamma_\mathrm{OFF}$, providing binary phase modulation with the same amplitude in two states. The metasurface design must therefore achieve $Z_\mathrm{S}=\eta$ when $Z_\mathrm{g}=\sqrt{Z_\mathrm{g,ON}\cdot Z_\mathrm{g,OFF}}$. For our samples we measure $Z_\mathrm{g,ON}=800 \;\Omega$, $Z_\mathrm{g,OFF}=4000 \;\Omega$. Therefore $\sqrt{Z_\mathrm{g,ON}\cdot Z_\mathrm{g,OFF}}= 1789 \;\Omega = 4.75 \eta$. This implies that the cell must be designed to scale down SLG's impedance to a factor 4.75 to be at the optimal working point.

To achieve this, we first chose a Salisbury screen configuration\cite{Salisbury}. This consists of a dielectric spacer on a reflective metallic layer\cite{TamagnoneMoldovanPoumirolEtAl2016b}. The spacer is a Si layer having thickness $t$:

\begin{equation}\label{seqn:salisbury}
t=\frac{\lambda_0}{4n}=\frac{c}{4nf_0}
\end{equation}

where $f_0 = 1$ THz is the design frequency, $\lambda_0$ is the corresponding free space wavelength, $c$ is the speed of light, and $n=\sqrt{11.7}$ is the Si refractive index. The purpose of this structure is to cancel the contribution of the reflective layer to the free space impedance, obtaining $Z_\mathrm{S}\simeq 0$ in absence of other structures on top of the spacer, as discussed in Ref.\citenum{TamagnoneMoldovanPoumirolEtAl2016b}. From Eq.\ref{seqn:salisbury} we get $t = 21.9$ $\mu$m. For our experiments we use $t = 20$ $\mu$m due to limitations in the available silicon on insulator (SOI) wafers.

The metallic structure in Fig.\ref{fig:SupplFigure4} is chosen to concentrate the impinging field on a SLG rectangular load over a broad-band, hence the choice of the bow-tie antenna element. SLG is prolonged on one side, to contact an additional Au bias line used to improve the connectivity of the column, so that the applied voltage is uniform even in case of cracks in one or more of the SLG loads. Voltage is applied to both ends of the column.

The cells ($20\times100$ $\mu$m$^2$) are smaller than half of the wavelength (300 $\mu$m at 1 THz). Each reflects the incident waves with a reflection coefficient that can be modulated applying different voltages to SLG. Numerical simulations and measurements of the reflection coefficient are in Fig.\ref{fig:Figure1}d-f. These measurements are performed by gating all the cells with the same voltage and then measuring the overall reflection coefficient $\Gamma$ of the surface, Fig.\ref{fig:Figure1}d. $\Gamma$ is a complex number describing both the amplitude and the phase of the reflected wave, with the phase delay normalized with respect to a reference mirror (Au deposited on the same chip directly on Al$_2$O$_3$). A THz fiber-coupled time domain system is used to measure $\Gamma$ (see Methods \ref{met:measurement}), focusing the incident beam on a small area of the array to avoid probing areas outside it. Note that, because of the subwavelength nature of the array, only one reflected beam exists, without diffraction.

The unit cell geometry is optimized so that, at the target design frequency of 1.05 THz, different $\sigma$ cause the reflection coefficient to vary from one value to its opposite, passing close to the total absorption condition ($\Gamma=0$). In this way, by switching the cell between these two states (ON and OFF) a local phase modulation of $\pi$ can be achieved. This is similar to the concept proposed in Refs.\citenum{sherrott2017experimental,kim2016vanadium} and demonstrated experimentally at microwave frequencies in Refs.\citenum{HumPerruisseau-Carrier2014,RodrigoJofrePerruisseau-Carrier2013}. The slight shift measured reflection coefficient with respect to the simulations visible in the figures is due to fabrication tolerances.

Beam steering can then be achieved by switching the cell state in the array so that a dynamical and reconfigurable phase hologram is created, obtaining a fully reconfigurable RA. We focus on beam steering and shaping in one plane. This allows for a simplification of the control network, whereby all cells belonging to the same column are connected together, and each column can be controlled by an individual voltage.

The far electric field radiation pattern obtained illuminating the array with a plane wave having electric field amplitude $E_0$ and angle of incidence $\THi$ (in our case fixed to 45$^\circ$) can be estimated based on the interference of discrete radiators\cite{Balanis2005}:

\begin{align}\label{eqn:mainformula}
E(\theta,r)=E_0\,g(r)\,\FC(\theta)\,\FA(\theta) = E_0\,g(r)\,\FC(\theta)
\nonumber\\ \,\sum_{n=1}^{N} \Gamma_n\,e^{\,jnk_0L(\sin\theta-\sin\THi)}
\end{align}

where $\theta$ is the deflection angle, $\FC(\theta)$ is the radiation pattern of a single isolated column, $\FA(\theta)$ is called \textit{array factor} \cite{Balanis2005}, $\Gamma_n$ is the reflection coefficient of the n-th cell, N is the total number of cells in the array, $k_0=2\pi/\lambda$ is the wavenumber, L is the cell width, r is the distance from the RA and $g(r)=r^{-1}exp(-jkr)$. The $\FC(\theta)$ factor is negligible here, as it does not show sharp variations in $\theta$ due to the sub-wavelength size of the unit cell. The summation (hence $\FA(\theta)$) is maximized when its elements are in phase. If a linear phase profile is created setting the $\Gamma_n$ elements such that $\Gamma_n=e^{jn\phi}$ then the maximum (hence the reflected beam direction) is obtained for $\theta=\arcsin\left(\sin\THi-\frac{\phi}{k_0 L}\right)$, which can be changed dynamically tuning the phase profile. It is possible to show (see Methods M3 for a full mathematical derivation) that this principle still holds if the reflection phase is quantized to just two values (0 and $\pi$) for all the elements, thus reducing the gradient to a periodic set of segments with phase 0 alternated with segments of phase $\pi$. The periodicity $P$ of the pattern expressed in terms of number of cells is then given by $P=2\pi/\phi$ and by the beam steering law:

\begin{equation}\label{eqn:steer}
\theta=\arcsin\left(\sin\THi-\frac{\lambda}{PL}\right)
\end{equation}

where $\lambda$ is the wavelength.

If $P$ is an even integer the pattern consists of a repetitions of a supercell of $P$ cells (with $P/2$ cells set to phase 0 and $P/2$ to $\pi$). Fig.\ref{fig:Figure2}a illustrates the case $P=4$. However, it is possible to generalize the pattern to odd and even fractional values of $P$ using a pattern generation technique described in Methods M4, thus achieving continuous beam steering.

Beam-steering with integer $P$ (from 4 to 8) is shown in Fig.\ref{fig:Figure2}b, for the voltage patterns in Fig.\ref{fig:Figure2}c. The angular steering range reaches 25$^\circ$. The beam is well-formed with the exception of small side lobes which appear for odd $P$, due to the technique used to emulate odd and fractional $P$ values. Fig.\ref{fig:Figure2}d plots the beam steering as a function of frequency, compared with the prediction of Eq.\ref{eqn:steer}, while Figs.\ref{fig:Figure2}e,f demonstrate the continuous beam-steering achieved with fractional $P$.

Our device can also reconfigure the beam shape. This is achieved by smoothly changing the local P from one extreme to the other of the array using a chirped pattern, as shown in Fig.\ref{fig:Figure2}k, thus having slightly different deflection angles across the array, emulating a parabolic profile. The device operates as a parabolic mirror with tunable curvature, which we use here to generate a wider beam, Fig.\ref{fig:Figure2}g. The same principle can be used to achieve tuneable focusing (limited here to one dimension).

Besides focusing and widening the beam, more complex operations can be performed. E.g., Fig.\ref{fig:Figure2}h plots the generation of a double beam by filling two halves of the array with patterns having different $P$ (changing abruptly in the middle of the array, as shown in the dual-beam pattern in Fig.\ref{fig:Figure2}k). Another important application is the ability to manipulate an impinging THz pulse at the time domain level. This is possible because the incident pulse reaches at different times each RA element. Therefore, the voltage pattern selected on the array is transferred to the time response of the system (within some limits due to the spectral response of each cell and to the total size of the illuminated portion of the array). Figs.\ref{fig:Figure2}i,j show that a periodic pattern with $P=6$ generates a sinusoid (of finite time duration due to the finite size of the array). The chirped pattern used for beam broadening gives a chirped sinusoid in the time response. Similar transformations can be achieved with more complex patterns.

We now consider the effect of shifting a periodic pattern (with $P = 2, 4, 6$) of a finite number of cells, and we verify that the corresponding time domain sinusoid is similarly de-phased. This is equivalent to the phase shift associated to a lateral translation of an optical grating\cite{lee2013displacement}, but the movement here is emulated by the reconfigurable control patterns. The experiment is illustrated in Fig.\ref{fig:Figure3}a and the measurements, better represented in the frequency domain, are plotted in Figs.\ref{fig:Figure3}b-d. These are the complex reflection coefficients for each of the aforementioned cases, and for each possible shift of the pattern. E.g., the $P=4$ patterns can be shifted in 4 ways, with shifts of 0, 1, 2, 3 cells, while shifting of 4 cells is identical to 0 and so on. Each cell shift corresponds to a phase delay of $2\pi/P$ regardless of the beam frequency. This scheme, here referred to as \textit{geometrical PSK} (phase shift keying\cite{blahut1987principles}), provides a way to perform a precise phase modulation on a wide band (60GHz at 1THz).

In summary, we reported a reconfigurable RA metasurface for terahertz waves using SLG. Beam steering, shaping and modulation were achieved. Our results demonstrate that graphene can be embedded in metasurfaces providing an unprecedented control and modulation capabilities for THz beams, with applications for adaptive optics, sensing and telecommunications. Our approach can be extended to mid infrared, and to two dimensional beam steering, by using individual cell control.

\section*{Acknowledgements}

We dedicate this work to the memory of Prof. Julien Perruisseau-Carrier. We thank Giancarlo Corradini, Cyrille Hibert, Julien Dorsaz, Joffrey Pernollet, Zdenek Benes, and the rest of EPFL CMi staff for the useful discussions. We acknowledge funding from the EU Graphene Flagship, the Swiss National Science Foundation (SNSF) grants 133583 and 168545, the Hasler Foundation (Project 11149), ERC Grant Hetero2D, EPSRC grant nos. EP/509 K01711X/1, EP/K017144/1, EP/N010345/1, EP/M507799/5101 and EP/ L016087/1.

\section*{Methods}

\subsection{Fabrication process flow}

\label{met:fab}

\begin{figure*}

\centerline{\includegraphics[width=180mm]{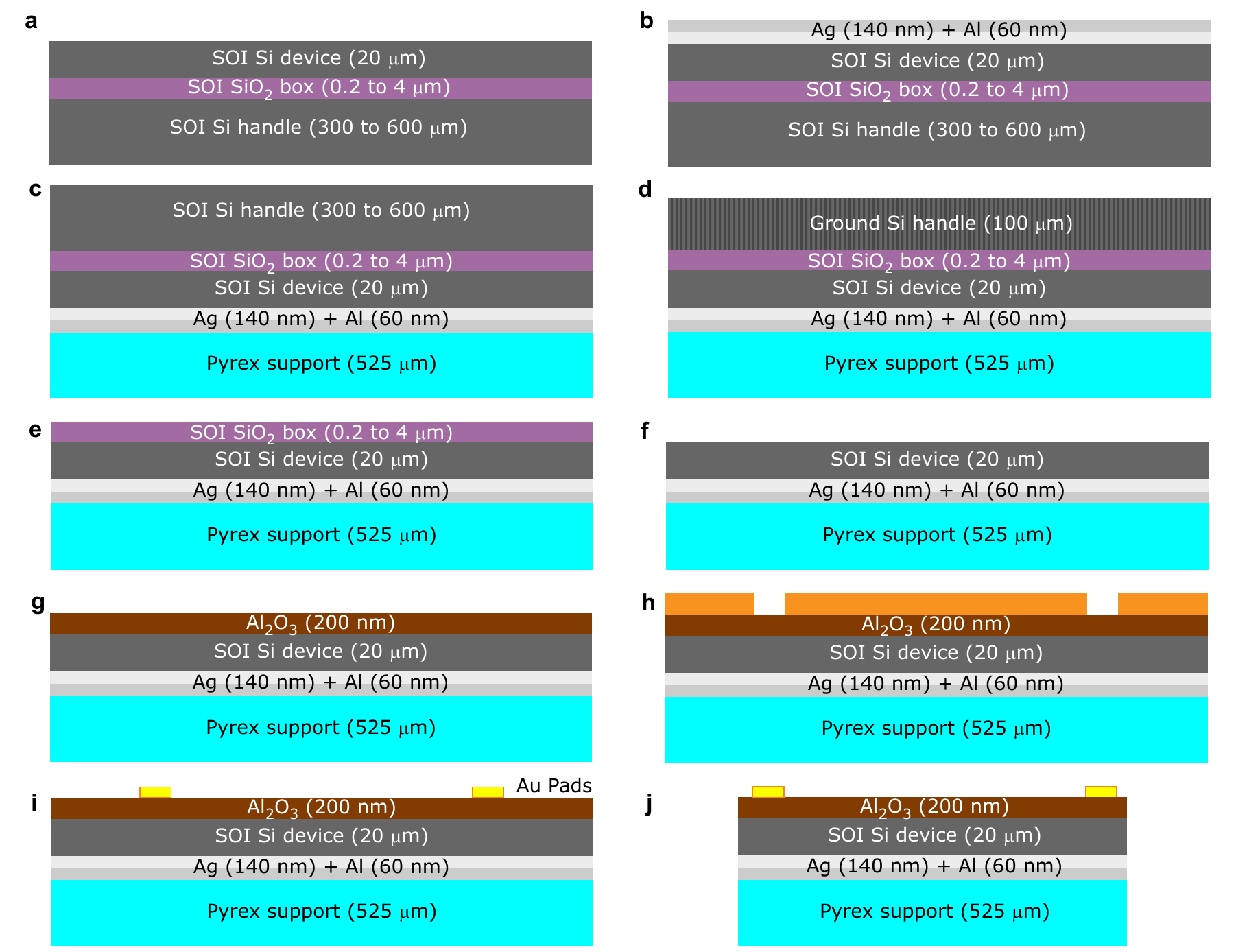}}

\caption{\label{fig:SupplFigure1}Fabrication process flow (part 1). \textbf{a}, Initial SOI wafer. \textbf{b}, E-beam evaporation of 140 nm Ag followed by 60 nm Al. \textbf{c}, Anodic bonding between the deposited Al layer and a support Pyrex wafer. This is cleaned in a hot piranha bath prior bonding. \textbf{d}, Grinding of the Si handle wafer down to 100 $\mu$m. \textbf{e}, Dry etching of the remaining Si. \textbf{f}, wet etching of SiO$_2$ box layer. \textbf{g}, ALD deposition of 200 nm of AlO$_2$. \textbf{h}, Optical lithography for the bonding pads: dual layer photo-resist spin coat, exposure and development. \textbf{i}, Oxygen plasma de-scum, e-beam evaporation of 100 nm Au pads with 5 nm Cr for adhesion and liftoff. \textbf{j}, Dicing.}

\end{figure*}

\begin{figure*}

\centerline{\includegraphics[width=180mm]{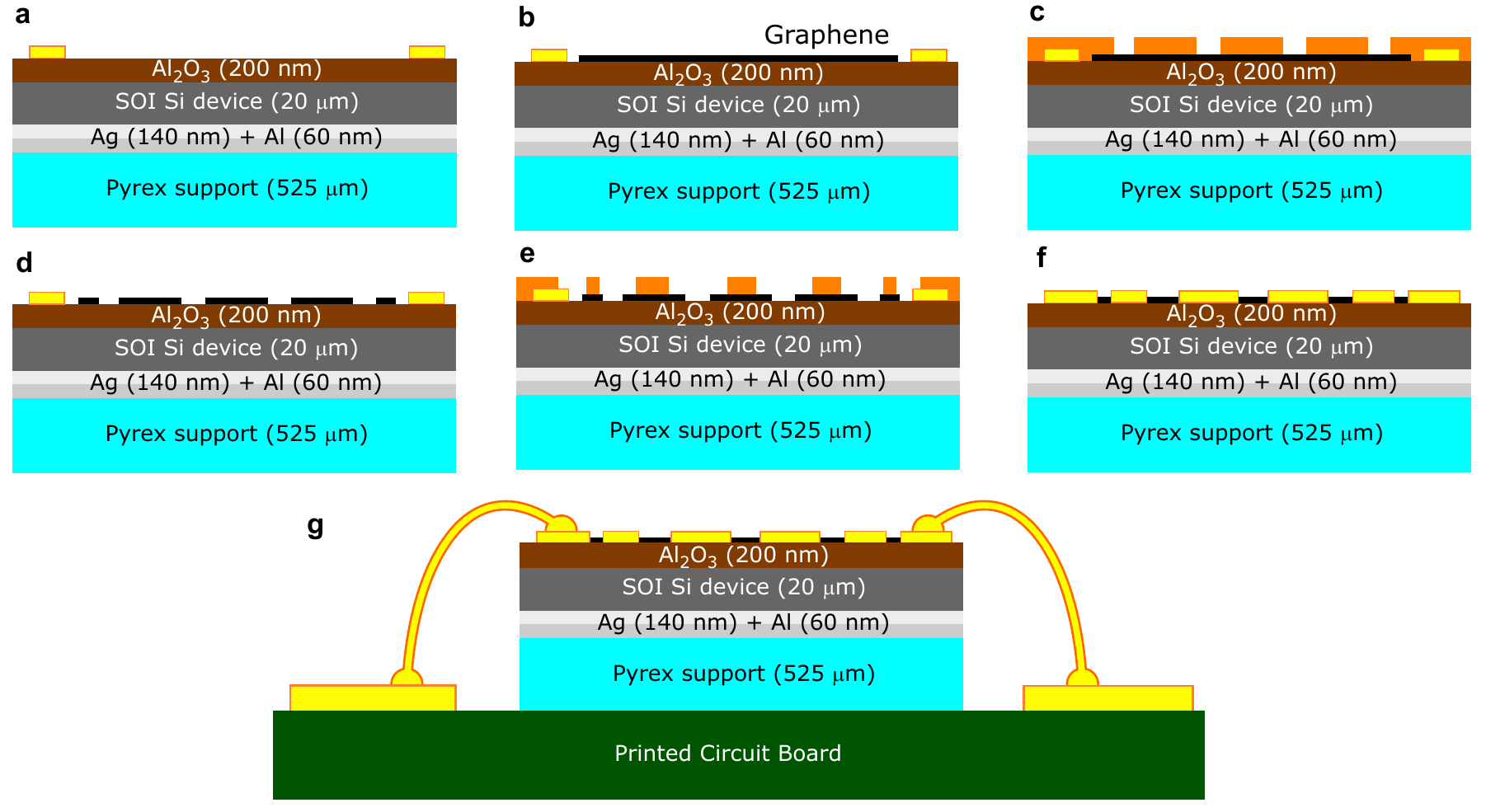}}

\caption{\label{fig:SupplFigure2}\textbf{a}, Chip as prepared in Fig.\ref{fig:SupplFigure1}. \textbf{b} SLG transfer. \textbf{c}, Lithography for SLG patterning; PMMA spin coat e-beam lithography and development. \textbf{d} SLG etching with oxygen plasma and PMMA stripping in hot acetone. \textbf{e}, Lithography for metallic antennas: MMA/PMMA spin coating, e-beam lithography and liftoff. \textbf{f}, Oxygen plasma de-scum, e-beam evaporation of 100 nm Au pads with 5 nm Cr for adhesion and liftoff. \textbf{g}, Mounting chip on PCB support and wire-bonding.}

\end{figure*}

\begin{figure*}
	
	\centerline{\includegraphics[width=120mm]{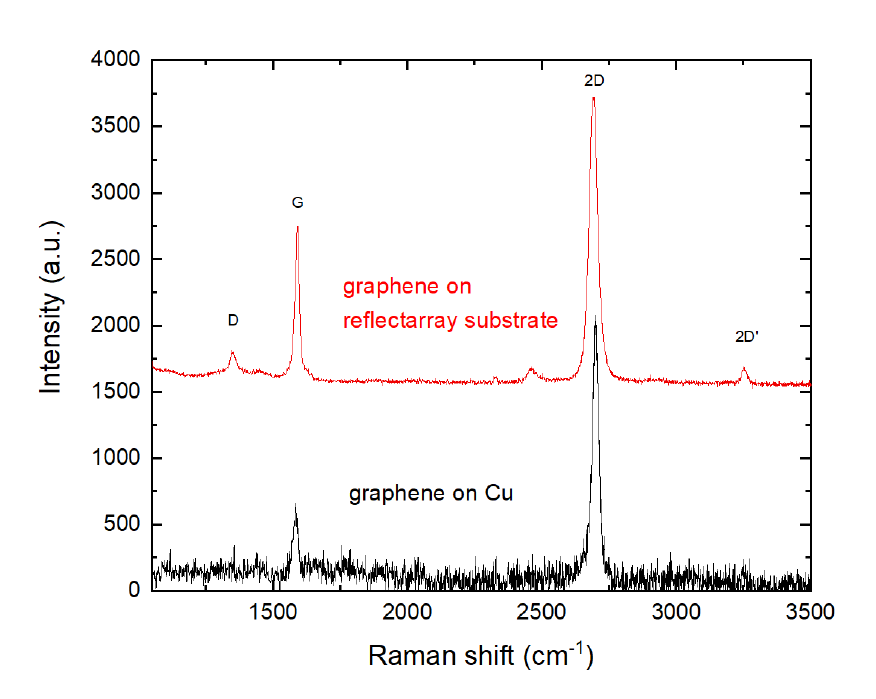}}
	
	\caption{\label{fig:SupplFigure9} Raman spectrum of the SLG before and after transfer.}
	
\end{figure*}

\begin{figure*}

\centerline{\includegraphics[width=180mm]{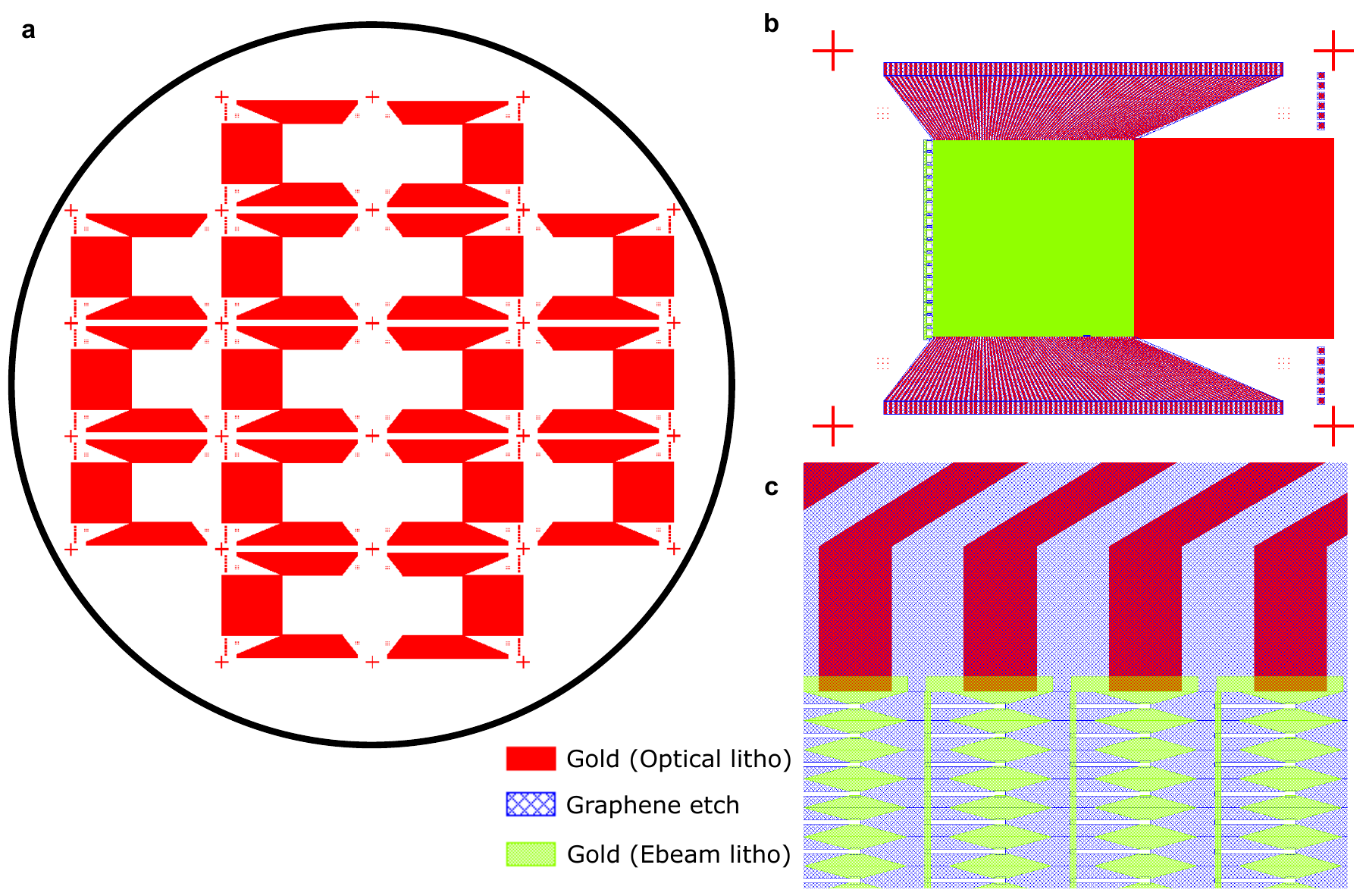}}

\caption{\label{fig:SupplFigure3}\textbf{a}, Full wafer, optical lithography mask for Au pads, reference mirrors and traces. \textbf{b}, Chip with all fabrication layers (Au optical lithography, SLG etch and Au e-beam lithography). \textbf{c} Magnification of the top part of the RA, showing RA columns.}

\end{figure*}

The starting point is a Si on insulator (SOI) wafer (produced by Ultrasil Corp.) having a device layer with the required characteristics for our dielectric spacer with 20 $\mu$m thickness and high resistivity $\sim$ 1 k$\Omega\cdot$cm, Fig.\ref{fig:SupplFigure1}a. Ag is deposited (e-beam evaporation) to create the reflective layer, followed by an Al layer (vacuum is not broken between the two depositions, Fig.\ref{fig:SupplFigure1}b). The Al coated face of the SOI wafer is then bonded via anodic bonding (Fig.\ref{fig:SupplFigure1}c) to a borosilicate glass wafer (Borofloat 33, very similar to Pyrex in composition\cite{borofloat}), acting as a support for the device layer, which is too fragile to be handled alone given its thickness. Bonding is performed at atmospheric pressure with a Suss Microtec SB6 tool immediately after evaporation. Glass wafers are also cleaned in a hot piranha bath immediately before bonding to remove organic impurities. A second borosilicate wafer is used below the one to be bonded, as a sacrificial substrate to collect the excess Na ions, thus preventing contamination. To prevent accidental bonding of the two borosilicate wafers, the sacrificial substrate is thinned using wafer grinding, and the non-polished surface is placed in contact with the borosilicate substrate to be bonded.

The aim of the following steps is to eliminate the SOI handle and box layer, to expose the device layer. This is done by first grinding the Si handle wafer down to 100 $\mu$m (Fig.\ref{fig:SupplFigure1}d). This is a mechanically aggressive process, therefore further thinning could damage the substrate or cause the failure of the bonding. The remaining Si is dry-etched using a fluorine-based chemistry, with a process having 200:1 selectivity with respect to SiO$_2$, Fig.\ref{fig:SupplFigure1}e. This ensures that the box layer survives the process, preserving the device layer as well. The box is then dissolved in HF 49\% (Fig.\ref{fig:SupplFigure1}f), selected over buffered oxide etch (BHF) because it etches faster SiO$_2$ \cite{williams1996etch} and, unlike BHF, does not attack Al \cite{williams1996etch}.

The gate oxide (200 nm Al$_2$O$_3$) is prepared using atomic layer deposition (ALD) on all the wafer, as shown in Fig.\ref{fig:SupplFigure1}g. Afterwards, dual layer photo-litography (LOR + AZ1512) is performed, followed by evaporation of 100 nm Au with an adhesion layer of 5 nm Cr and liftoff (Fig.\ref{fig:SupplFigure1}h,i). During this step, the reference mirrors (one for each chip), bonding pads, and dicing markers are defined on the full wafer (Fig.\ref{fig:SupplFigure3}a). Oxygen plasma de-scum is performed prior to the evaporation to ensure maximum adhesion, important for the subsequent wire-bonding step. The wafer is then diced (Fig.\ref{fig:SupplFigure1}j) using an automatic dicing saw (Disco DAD-321). During the dicing process, the wafer is protected by a photo-resist layer, then stripped in remover on each chip, lifting the dicing residues.

SLG is grown on Cu foil (99.8\% purity) by chemical vapor deposition (CVD) on a tube furnace as for Ref.\citenum{li}. The Cu foil is annealed in H$_2$ (flow 20 sccm) at 1000 $^\circ$C for 30 min. After annealing, CH$_4$ (flow 5 sccm) is introduced for 30 min while keeping the temperature at 1000 $^\circ$C, leading to the growth of SLG. This is then transferred onto the Al$_2$O$_3$/Si/Ag/Pyrex by wet transfer (Figure \ref{fig:SupplFigure2}a,b)\cite{bae}, where polymethyl methracrylate (PMMA) is used as a sacrificial layer to support SLG during Cu etching in ammonium persulfate\cite{bae}. After transfer, PMMA is dissolved in acetone. Raman spectroscopy is used to monitor the sample quality throughout the process by using a Renishaw inVia spectrometer equipped with 100X objective and a 2400 groves/mm grating at 514.5 nm. Representative Raman spectra of SLG placed onto the Al$_2$O$_3$/Si/Ag/Pyrex substrate are shown in Fig.\ref{fig:SupplFigure9}. The spectrum of graphene on Cu shows not significant D peak, indicating a negligible defect density \cite{NN2013}. After transfer, the position of G peak is 1590 cm$^{-1}$ and its full width at half maximum is 17 cm$^{-1}$, the position of 2D peak is 2692 cm$^{-1}$, while the ratio of the 2D to G peaks intensities, I(2D)/I(G), is $\sim$1.84 and the ratio of their areas, A(2D)/A(G), is $\sim$4, indicating a Fermi level $\sim$0.2-0.4eV and a charge carrier concentration $\sim$10$^{12}$ cm$^{-2}$ \cite{Das}. The D peak is present in the spectrum of the transferred SLG, suggesting that some defects have been introduced during the process. From I(D)/I(G) $\sim$0.13 and given the Fermi level, we can estimate a defect density $\sim$7x10$^{10}$ cm$^{-2}$ \cite{Bruna2014,cancado}.

SLG is then e-beam patterned using PMMA resist followed by oxygen plasma and stripping in acetone at 45 $^\circ$C, Fig.\ref{fig:SupplFigure2}c,d). Apart from patterning SLG in the RA, during this process all the bonding and traces are also exposed to the oxygen plasma to ensure that no SLG remains on them, to avoid short circuits. Subsequently, a new e-beam lithography (MMA + PMMA) is performed to define the metallic antennas via evaporation and liftoff in acetone, Figs. \ref{fig:SupplFigure2}e,f, \ref{fig:SupplFigure3}a,b. Finally the chip is glued to the PCB substrate and all the columns are connected via wire-bonding to the PCB traces, Fig.\ref{fig:SupplFigure2}g). The ground plane is contacted laterally with Ag paint.

\subsection{Measurement setup and post processing}

\label{met:measurement}

\begin{figure*}

\centerline{\includegraphics[width=180mm]{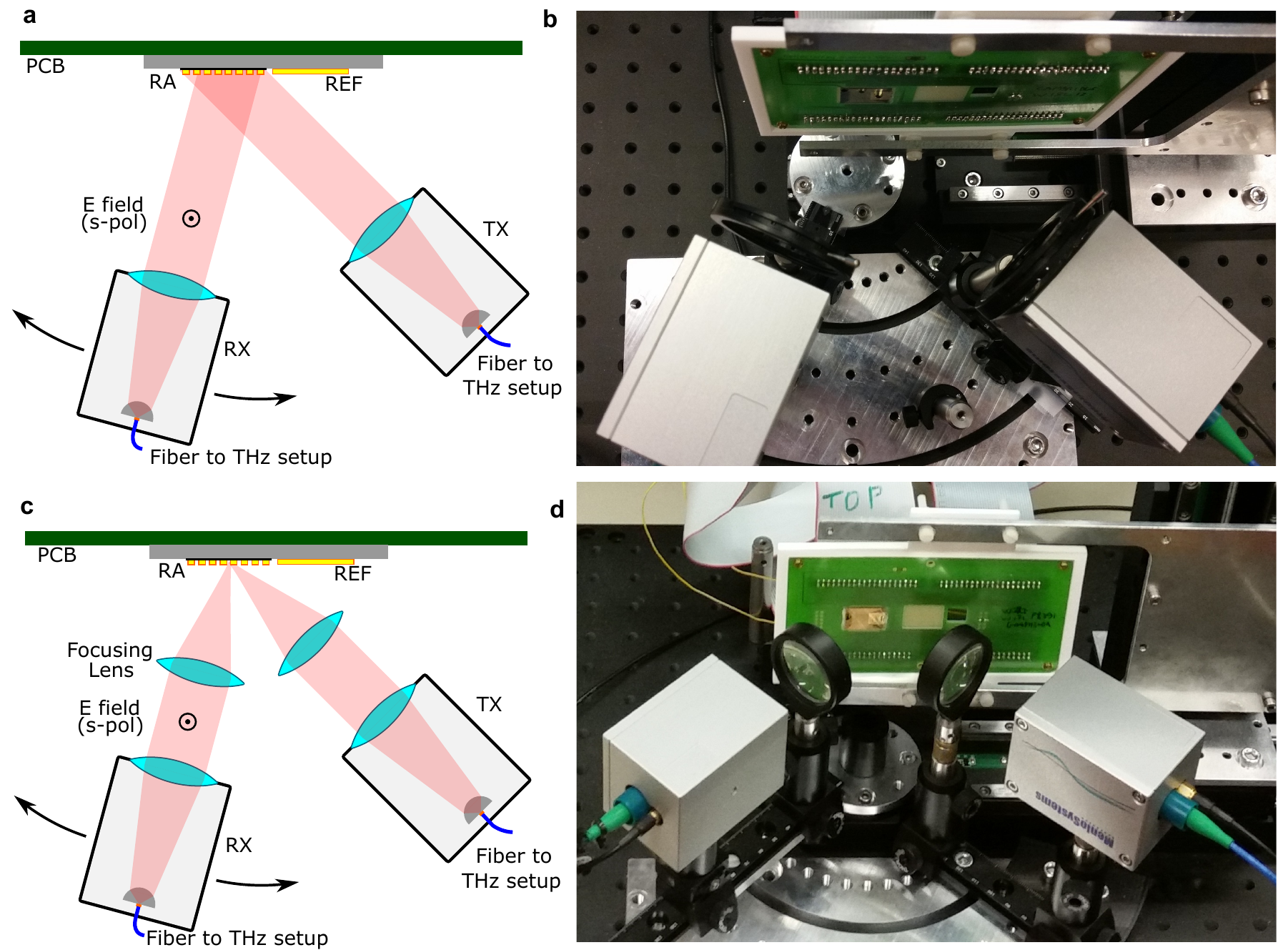}}

\caption{\label{fig:SupplFigure6} \textbf{a-b}, Collimated beam measurement, schematics and picture. TX, THz emitter; RX, detector. RA, reflectarray; REF, reference mirror. \textbf{c-d}, Focused beam measurement, schematics and picture. The focusing lens of the detector rotates together with the detector itself.}

\end{figure*}

The RA is characterized using a commercial fiber coupled THz time domain setup (by Menlo Systems, model TERA K15 mark II). The THz emitter is mounted at 45 degrees of incidence with respect to the RA, while the detector is placed on a motorized rotation stage. The system is first aligned in transmission to maximize the THz signal intensity, and subsequently in reflection, using a reference mirror mounted on the sample holder. This is mounted on a translation stage (motorized XY linear stages plus manual Z stage), to automatically alternate between sample and reference mirror. All measurements are normalized with respect to the reference mirror, created on the same chip of the RA during the optical lithography process on the full wafer.

Two different measurement modes are used, Figure \ref{fig:SupplFigure6}:

\begin{itemize}

	\item \textbf{Collimated} The beam is collimated and impinges on a large area of the sample ($\sim$1 cm$^2$). The reduced spread of the angular components of the beam (when decomposed into a superposition of plane waves) ensures precise measurement of angles and radiation patterns, but part of the beam interacts with the area around the RA.

	\item \textbf{Focused}: The beam is focused by an additional pair of lenses so that it impinges completely inside the array. However, this requires larger spread of the angular components, hence this mode is not accurate for angles and radiation pattern measurement. Instead, it is used to measure the reflection coefficient and the efficiency of the array.

\end{itemize}

The measurements in Fig.\ref{fig:Figure1} are performed in the \textit{focused} mode, while those in Figs.\ref{fig:Figure2},\ref{fig:Figure3} use the \textit{collimated} mode. The latter is to be preferred to characterize the geometric phase shift keying (G-PSK, Fig.\ref{fig:Figure3}), since precise phase modulation relies on the interaction between the beam and all of the columns of the RA, which can be illuminated completely only in the collimated mode.

The drawback of the collimated mode is the interaction of the beam with areas outside the RA metasurface. This can be addressed with the following post-processing method. For radiation pattern characterization, two measurements are performed, one with the chosen control sequence, and the other with the opposite (logical NOT) control sequence. In this way, the steered beam will have opposite phase in the two cases (see Method \ref{met:radpattheo}) and the radiation pattern can be extracted by subtracting (frequency by frequency and angle by angle) these two measurements. Any contribution from the area outside the array is canceled by the subtraction. For the G-PSK case, the same is accomplished by subtracting from all the signals the average in the complex plane (frequency by frequency). An overall phase factor $e^{-j(\omega\tau+\phi)}$ is removed from all the traces. A unique value of the delay $\tau$ is used for all the symbols in each G-PSK measurement. This is done to remove the free-space phase delay of the measured beam, due to slight differences in the paths when measuring the array and the reference mirror. Similarly, the removed phase factor $\phi$ is unique for all the traces, and it is used to align the symbols to the real and imaginary axes of the complex plane.

\subsection{Radiation pattern theory}

\label{met:radpattheo}

The array geometry does not depend on the $y$ direction since the array has a periodicity smaller than half wavelength in that direction, independently of the control pattern. We assume that the incident wave is propagating in the $xz$ plane (i.e. $k_y=0$). In the low cell-to-cell coupling approximation\cite{HumPerruisseau-Carrier2014}, the electric field of an antenna array in the $x$ direction (assuming equidistant elements separated by $L$) in far field conditions is given by\cite{HumPerruisseau-Carrier2014, Balanis2005}:

\begin{align}\label{seqn:mainformula}
E(\theta,r) = g(r)\sum_{n=1}^{N} w_n\FC(\theta)\,e^{\,jnk_0L\sin\theta} = g(r)\FC(\theta)\nonumber\\ \sum_{n=1}^{N}w_n\,e^{\,jnk_0L\sin\theta}
\end{align}

where $\FC$ is the single cell radiation pattern, $k_0$ is the wavenumber, $w_n$ is the amplitude associated to the n-th element. For a RA we can write $w_n$ as the product of the incident field at the element position times a reflection coefficient:

\begin{equation}\label{seqn:wn}
w_n=\Gamma_n\, \Ei(x=nL,z=0)
\end{equation}

where we assume for simplicity and without loss of generality that the RA is in the $z=0$ plane. The electric field $\Ei$ of an incident wave (with incident angle $\THi$ with respect to the normal) is given by:

\begin{equation}\label{seqn:einc}
\Ei(x,z)=E_0\,e^{-jk_0(x\sin\THi+z\cos\THi)}
\end{equation}

Combining Eqs.\ref{seqn:einc},\ref{seqn:wn},\ref{seqn:mainformula} we get:

\begin{align}\label{seqn:mainformula2}
E(\theta) = E_0 \, g(r) \FC(\theta) \sum_{n=1}^{N}\Gamma_n\,e^{\,jnk_0L(\sin\theta-\sin\THi)}=\nonumber\\ E_0 \, g(r) \FC(\theta)\FA(\theta)
\end{align}

where we define the \textit{array factor} $\FA(\theta)$ as:

\begin{equation}\label{seqn:af}
\FA(\theta) = \sum_{n=1}^{N}\Gamma_n\,e^{\,jnk_0L(\sin\theta-\sin\THi)}
\end{equation}

We notice that, if all the reflection coefficients are phase-modulated of $\pi$ (thus reversing their sign), the total phase of the scattered field will also be out of phase of $\pi$, which is used to suppress the background in our measurements. If all the cells have the same $\Gamma$, then:

\begin{equation}\label{seqn:uniformarray}
E(\theta) = E_0\, g(r)\, \FC(\theta)\,\Gamma\sum_{n=1}^{N}\,e^{\,jnk_0L(\sin\theta-\sin\THi)}
\end{equation}

A maximum in the reflection is obtained for $\theta=\THi$ (which is the direction of the specular reflection) where all the contributions of the summation add in phase. More generally, a maximum is obtained if:

\begin{equation}\label{seqn:refl}
k_0 L(\sin\theta-\sin\THi)=2\pi l \qquad l\in \mathbb{Z}
\end{equation}

for $\lambda=2\pi/k_0$:

\begin{equation}\label{seqn:refl2}
\sin\theta-\sin\THi= l \frac{\lambda}{L} \qquad l\in \mathbb{Z}
\end{equation}

Because $\lambda/L=3$ in our case, only the specular reflection $l=0$ satisfies this condition, for any $\THi$.

Super-cells can be implemented by creating periodic patterns of reflection coefficients, fulfilling the condition $\Gamma_n=\Gamma_{n+P}$, where $P$ is a positive integer number of cells in the super-cell. If a super-cell with periodicity $P$ is implemented, then the array factor can be rewritten, by splitting the summation in two levels: an external sum over all the super-cells, and an internal one over the cells in a super-cell:

\begin{align}\label{seqn:supercelldecomp}
\FA(\theta) =
\sum_{n=0}^{N/P}\sum_{m=1}^{P}\Gamma_m\,e^{\,j(nP+m)k_0L(\sin\theta-\sin\THi)}=\nonumber\\
\sum_{m=1}^{P}\Gamma_m\,e^{\,jmk_0L(\sin\theta-\sin\THi)}\sum_{n=0}^{N/P}\,e^{\,jnPk_0L(\sin\theta-\sin\THi)}
\end{align}

We can then define the \textit{supercell factor} $\FSC$:

\begin{equation}\label{seqn:supercell}
\FSC(\theta) =\sum_{m=1}^{P}\Gamma_m\,e^{\,jmk_0L(\sin\theta-\sin\THi)}
\end{equation}

and the \textit{superarray factor} $\FSA$:

\begin{equation}\label{seqn:superarray}
\FSA(\theta) =\sum_{n=0}^{N/P}\,e^{\,jnPk_0L(\sin\theta-\sin\THi)}
\end{equation}

and use them to decompose the array factor as $\FA(\theta)=\FSC(\theta) \FSA(\theta)$. The final expression for the electric far field is then:

\begin{align}\label{seqn:mainformula3}
E(\theta) \;=\; E_0 \, g(r) f(\theta) \;=\; E_0 \, g(r) \FC(\theta)\FA(\theta) \;=\nonumber\\ \; E_0 \, g(r) \FC(\theta) \FSC(\theta) \FSA(\theta)
\end{align}

The angular part of the radiation pattern $f(\theta)$ is decomposed in three factors (ordered here from the most directive to the least):

\begin{itemize}

	\item $\FSA$, associated to the super-array, identifies a set of possible directions where light can be scattered, and behaves as a diffraction grating.

	\item $\FSC$, associated to the super-cell, gives different weights to the possible diffraction orders accordingly to the phase of the cells in the supercell.

	\item $\FC$, associated to the cell, slowly varying with $\theta$ in the subwavelength case and does not affect significantly the final pattern.

\end{itemize}

The directions of diffracted beams launched by the array is then given by $\FSA(\theta)$, and for each beam the following must be satisfied:

\begin{equation}\label{seqn:grating}
\sin\theta-\sin\THi= l \frac{\lambda}{PL} \qquad l\in \mathbb{Z}
\end{equation}

Evaluating $\FSC$ for each of the diffracted beams:

\begin{align}\label{seqn:fscl}
\FSC(l)=\sum_{m=1}^{P}\Gamma_m\,e^{\,jmk_0L(\sin\theta-\sin\THi)}=\nonumber\\ \sum_{m=1}^{P}\Gamma_m\,e^{\,j 2 \pi m  l /P} \qquad l\in \mathbb{Z}
\end{align}

\begin{figure*}

\centerline{\includegraphics[width=180mm]{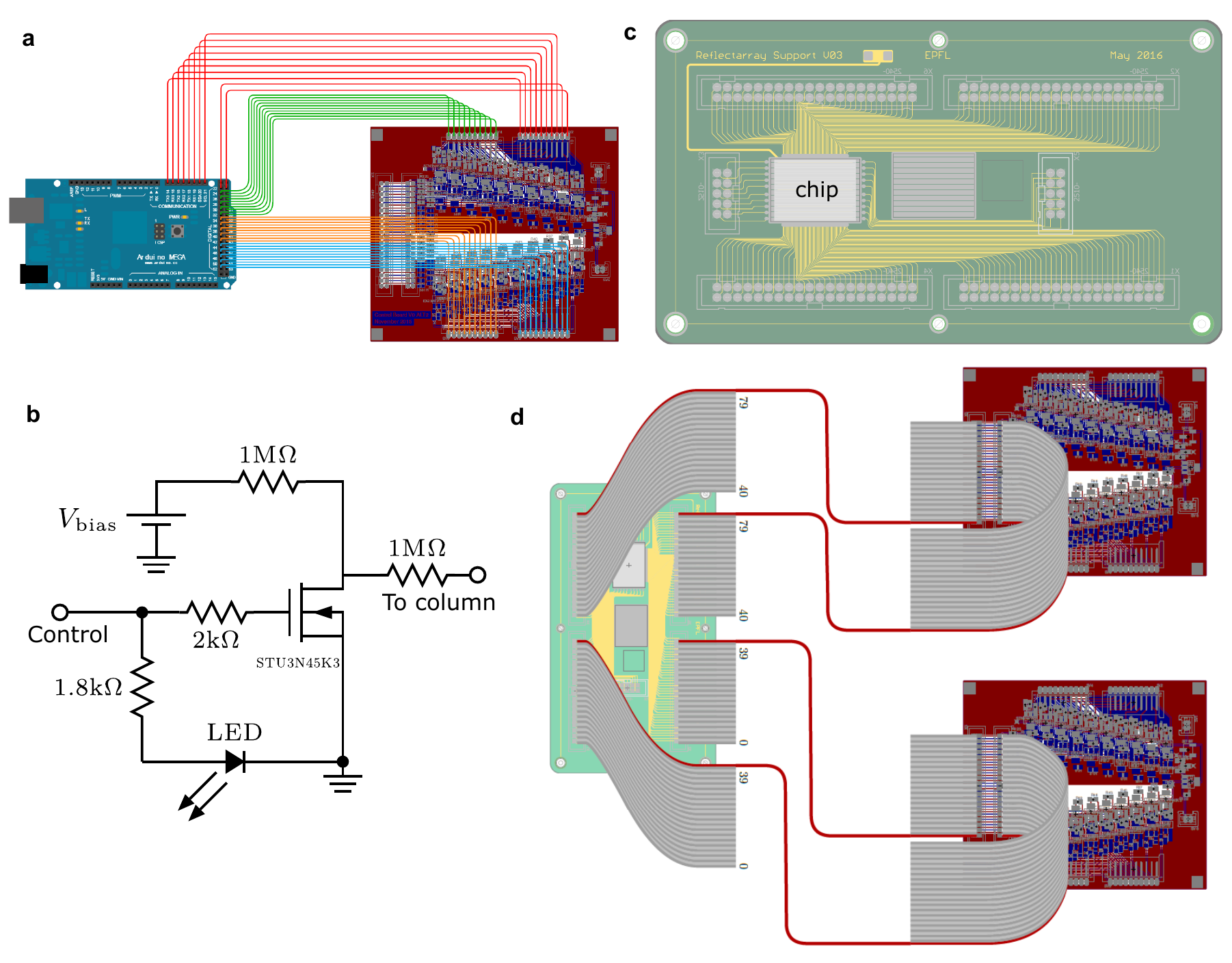}}

\caption{\label{fig:SupplFigure5} \textbf{a}, Arduino unit connected to a PCB driver. \textbf{b} Transistor stage used to drive the columns (there are 40 of these in each PCB driver). \textbf{c} PCB support, on which the RA is glued and wirebonded. \textbf{d} Connections between PCB support and drivers.}

\end{figure*}

In our case $\THi=45^\circ$ and we are operating at 1 THz. Also, let us consider the simplest case, with $N$ even integer and with a supercell formed by $N/2$ cells with reflection phase $0^\circ$ followed by $N/2$ cells with reflection phase $180^\circ$. $l=-1$ represents the steered beam of interest. The specular reflection ($l=0$) is suppressed because $\FSC$ vanishes for $\theta=\THi$. This is due to the fact that in our super-cell half of the cells have a reflection coefficient $\Gamma$ and the remaining ones $-\Gamma$, so the total sum is zero.

Then evaluating the summation for the considered supercell:

\begin{align}\label{seqn:evenN}
\FSC(l)=\sum_{m=1}^{P}\Gamma_m\,e^{\,j 2 \pi m  l /P}=\nonumber\\
\Gamma \left(\sum_{m=1}^{P/2}\,e^{\,j 2 \pi m  l /P}-\sum_{m=P/2+1}^{P}\,e^{\,j 2 \pi m  l /P}\right)
\end{align}

The same cancelation holds for any even value of $l$. Any beam for odd values of $l$ different from 1 and -1 is also strongly attenuated. For all our choices of period ($P$ between 4 and 8) the $l=1$ beam does not exist, since no real $\theta$ solves Eq.\ref{seqn:grating} for the chosen beam wavelength and incident angle. Imperfections in the array may still cause smaller side lobes, i.e. unwanted beams for $l\neq-1$. This is especially true for odd and fractional values of $P$, where the analysis of Eq.\ref{seqn:evenN} does not apply rigorously (though it still describes qualitatively the situation).

The geometric PSK operation can be understood inspecting $\FSC$ for $l=-1$, and noting that a translation of any supercell pattern described as $\Gamma_m \leftarrow  \Gamma_{(m+1) \mathrm{mod} P}$ creates a phase shift $2\pi/P$ (for an infinite array approximation):

\begin{equation}\label{seqn:QPSK1}
\FSC(l=-1)=
\sum_{m=1}^{P}\Gamma_m\,e^{\,-j 2 \pi m  /P}
\end{equation}

\begin{align}\label{seqn:QPSK2}
\sum_{m=1}^{P}\Gamma_{(m+1)\mathrm{mod} P}\,e^{\,-j 2 \pi m  /P}=
\sum_{r=0}^{P-1}\Gamma_{r}\,e^{\,-j 2 \pi(r-1)/P}=\nonumber\\
e^{j2\pi/P}\sum_{r=1}^{P}\Gamma_{r}\,e^{\,-j 2 \pi r/P}
\end{align}

The software Ansys HFSS is used to compute and optimize the reflection coefficient of the cells. In particular, the effect of the metallic bias lines on the cell impedance is reduced thanks to the optimization. Simulations are performed using the Floquet boundary conditions (equivalent to Bloch periodic boundary conditions, see for example\cite{TamagnoneFallahiMosigEtAl2014}), with an incident angle $45^\circ$. The optimized bias line width is 2 $\mu$m, so that its inductance per unit length is sufficient to reduce its effect on the structure.

\subsection{Control unit and list of control sequences}

\label{met:control}

The RA is glued on a support PCB substrate and each column is wirebonded to a PCB trace. Both ends of each column are connected to a high voltage transistor stage, and all the stages are controlled by two Arduino units. Fig.\ref{fig:SupplFigure5}a is an Arduino unit connected to a PCB driver with 40 transistor stages. A similar unit controls the remaining 40 columns. Fig.\ref{fig:SupplFigure5}b shows a circuit schematic of the high voltage stage, while Fig.\ref{fig:SupplFigure5}c illustrates the support PCB substrate. Fig.\ref{fig:SupplFigure5}d has two PCB drivers connected to the support PCB. The high voltage stage can switch on and off the voltage of each column, reaching values close to the supply voltage in one case and close to ground in the other. The needed positive and negative gate voltages are achieved by applying an intermediate voltage on the retroreflector of the RA, so that the voltage difference is negative in one state and positive in the other one. The total voltage is set via a high voltage DC generator, and the retroreflector voltage is controlled by a potentiometer.

Tables \ref{tab:SupplTable1}, \ref{tab:SupplTable2}, \ref{tab:SupplTable3}, \ref{tab:SupplTable4} show the control sequences used in each of our experiments. These are programmed into the Arduino module by the control computer. The latter can control the Arduino modules, the rotary stage, the XY stage, the high voltage generator and the terahertz setup, so that the measurements are completely automatized.

ON and OFF states (or equivalently 1 and 0) correspond to the following voltages between SLG and retroreflector: $V_\mathrm{ON}=26V$, $V_\mathrm{OFF}=-44 V$. The sequences are generated by discretizing sinusoids and chirped sinusoids with different periods, which leads to quasi-periodic signals for odd and fractional P values.

\begin{table*}

		\small
		\fontsize{9}{10}\selectfont

		\caption{List of control sequences for coarse beam steering}

		\label{tab:SupplTable1}

		\begin{tabular}{|l|l|} 

			\hline

			Sequence name & Sequence \\

			\hline

			$\phantom{\lnot}$ All OFF \rule{0pt}{3ex} & 00000000 00000000 00000000 00000000 00000000 00000000 00000000 00000000 00000000 00000000 \\

			$\phantom{\lnot}$ All ON & 11111111 11111111 11111111 11111111 11111111 11111111 11111111 11111111 11111111 11111111 \\

			$\phantom{\lnot}$ Period 4 \rule{0pt}{3ex} & 10011001 10011001 10011001 10011001 10011001 10011001 10011001 10011001 10011001 10011001 \\

			$\lnot$ Period 4 & 01100110 01100110 01100110 01100110 01100110 01100110 01100110 01100110 01100110 01100110 \\

			$\phantom{\lnot}$ Period 5 & 11001100 01110011 00011100 11000111 00110001 11001100 01110011 00011100 11000111 00110001 \\

			$\lnot$ Period 5 & 00110011 10001100 11100011 00111000 11001110 00110011 10001100 11100011 00111000 11001110 \\

			$\phantom{\lnot}$ Period 6 & 00011100 01110001 11000111 00011100 01110001 11000111 00011100 01110001 11000111 00011100 \\

			$\lnot$ Period 6 & 11100011 10001110 00111000 11100011 10001110 00111000 11100011 10001110 00111000 11100011 \\

			$\phantom{\lnot}$ Period 7 & 10001110 00011110 00111000 01111000 11100001 11100011 10000111 10001110 00011110 00111000 \\

			$\lnot$ Period 7 & 01110001 11100001 11000111 10000111 00011110 00011100 01111000 01110001 11100001 11000111 \\

			$\phantom{\lnot}$ Period 8 & 11100001 11100001 11100001 11100001 11100001 11100001 11100001 11100001 11100001 11100001 \\

			$\lnot$ Period 8 & 00011110 00011110 00011110 00011110 00011110 00011110 00011110 00011110 00011110 00011110 \\

			\hline

		\end{tabular}

		\caption{List of control sequences for fine beam steering}

		\label{tab:SupplTable2}

		\begin{tabular}{|l|l|} 

			\hline

			Sequence name & Sequence \\

			\hline

			$\phantom{\lnot}$ All OFF \rule{0pt}{3ex} & 00000000 00000000 00000000 00000000 00000000 00000000 00000000 00000000 00000000 00000000 \\

			$\phantom{\lnot}$ All ON & 11111111 11111111 11111111 11111111 11111111 11111111 11111111 11111111 11111111 11111111 \\

			$\phantom{\lnot}$ Period 5.5 \rule{0pt}{3ex}& 01110001 11001110 00111001 11000111 00111000 11100011 00011100 01100011 10001100 01110001 \\

			$\lnot$ Period 5.5 & 10001110 00110001 11000110 00111000 11000111 00011100 11100011 10011100 01110011 10001110 \\

			$\phantom{\lnot}$ Period 5.75 & 11100111 00011100 01110001 11001110 00111000 11100011 10001100 01110001 11000111 00011000 \\

			$\lnot$ Period 5.75 & 00011000 11100011 10001110 00110001 11000111 00011100 01110011 10001110 00111000 11100111 \\

			$\phantom{\lnot}$ Period 6 & 10001110 00111000 11100011 10001110 00111000 11100011 10001110 00111000 11100011 10001110 \\

			$\lnot$ Period 6 & 01110001 11000111 00011100 01110001 11000111 00011100 01110001 11000111 00011100 01110001 \\

			$\phantom{\lnot}$ Period 6.25 & 00111100 01110001 11000111 00011110 00111000 11100011 10000111 00011100 01110001 11000011 \\

			$\lnot$ Period 6.25 & 11000011 10001110 00111000 11100001 11000111 00011100 01111000 11100011 10001110 00111100 \\

			$\phantom{\lnot}$ Period 6.5 & 01110001 11100011 10001111 00011100 01111000 11100001 11000111 00001110 00111000 01110001 \\

			$\lnot$ Period 6.5 & 10001110 00011100 01110000 11100011 10000111 00011110 00111000 11110001 11000111 10001110 \\

			\hline

		\end{tabular}

		\caption{List of control sequences for beam shaping}

		\label{tab:SupplTable3}

		\begin{tabular}{|l|l|} 

			\hline

			Sequence name & Sequence \\

			\hline

			$\phantom{\lnot}$ All OFF \rule{0pt}{3ex} & 00000000 00000000 00000000 00000000 00000000 00000000 00000000 00000000 00000000 00000000 \\

			$\phantom{\lnot}$ All ON & 11111111 11111111 11111111 11111111 11111111 11111111 11111111 11111111 11111111 11111111 \\

			$\phantom{\lnot}$ Period 6\rule{0pt}{3ex} & 10001110 00111000 11100011 10001110 00111000 11100011 10001110 00111000 11100011 10001110 \\

			$\lnot$ Period 6 & 01110001 11000111 00011100 01110001 11000111 00011100 01110001 11000111 00011100 01110001 \\

			$\phantom{\lnot}$ Chirp 1 & 01110011 10001100 01110001 10001110 00111000 11100011 10001110 00011100 01111000 11100001 \\

			$\lnot$ Chirp 1 & 10001100 01110011 10001110 01110001 11000111 00011100 01110001 11100011 10000111 00011110 \\

			$\phantom{\lnot}$ Chirp 2 & 00111001 11001110 01110001 11001110 00111000 11100011 10001111 00011100 00111000 01111000 \\

			$\lnot$ Chirp 2 & 11000110 00110001 10001110 00110001 11000111 00011100 01110000 11100011 11000111 10000111 \\

			$\phantom{\lnot}$ Dual beam & 11001100 11001100 11001100 11001100 11001100 11110000 11110000 11110000 11110000 11110000 \\

			$\lnot$ Dual beam & 00110011 00110011 00110011 00110011 00110011 00001111 00001111 00001111 00001111 00001111 \\

			\hline

		\end{tabular}

		\caption{List of control sequences for geometric PSK}

		\label{tab:SupplTable4}

		\begin{tabular}{|l|l|} 

			\hline

			Sequence name & Sequence \\

			\hline

			All OFF \rule{0pt}{3ex} & 00000000 00000000 00000000 00000000 00000000 00000000 00000000 00000000 00000000 00000000 \\

			All ON & 11111111 11111111 11111111 11111111 11111111 11111111 11111111 11111111 11111111 11111111 \\

			4-GPSK Symbol 0 \rule{0pt}{3ex}& 10011001 10011001 10011001 10011001 10011001 10011001 10011001 10011001 10011001 10011001 \\

			4-GPSK Symbol 1 & 00110011 00110011 00110011 00110011 00110011 00110011 00110011 00110011 00110011 00110011 \\

			4-GPSK Symbol 2 & 01100110 01100110 01100110 01100110 01100110 01100110 01100110 01100110 01100110 01100110 \\

			4-GPSK Symbol 3 & 11001100 11001100 11001100 11001100 11001100 11001100 11001100 11001100 11001100 11001100 \\

			6-GPSK Symbol 0 \rule{0pt}{3ex}& 00011100 01110001 11000111 00011100 01110001 11000111 00011100 01110001 11000111 00011100 \\

			6-GPSK Symbol 1 & 00111000 11100011 10001110 00111000 11100011 10001110 00111000 11100011 10001110 00111000 \\

			6-GPSK Symbol 2 & 01110001 11000111 00011100 01110001 11000111 00011100 01110001 11000111 00011100 01110001 \\

			6-GPSK Symbol 3 & 11100011 10001110 00111000 11100011 10001110 00111000 11100011 10001110 00111000 11100011 \\

			6-GPSK Symbol 4 & 11000111 00011100 01110001 11000111 00011100 01110001 11000111 00011100 01110001 11000111 \\

			6-GPSK Symbol 5 & 10001110 00111000 11100011 10001110 00111000 11100011 10001110 00111000 11100011 10001110 \\

			8-GPSK Symbol 0 \rule{0pt}{3ex}& 11100001 11100001 11100001 11100001 11100001 11100001 11100001 11100001 11100001 11100001 \\

			8-GPSK Symbol 1 & 11000011 11000011 11000011 11000011 11000011 11000011 11000011 11000011 11000011 11000011 \\

			8-GPSK Symbol 2 & 10000111 10000111 10000111 10000111 10000111 10000111 10000111 10000111 10000111 10000111 \\

			8-GPSK Symbol 3 & 00001111 00001111 00001111 00001111 00001111 00001111 00001111 00001111 00001111 00001111 \\

			8-GPSK Symbol 4 & 00011110 00011110 00011110 00011110 00011110 00011110 00011110 00011110 00011110 00011110 \\

			8-GPSK Symbol 5 & 00111100 00111100 00111100 00111100 00111100 00111100 00111100 00111100 00111100 00111100 \\

			8-GPSK Symbol 6 & 01111000 01111000 01111000 01111000 01111000 01111000 01111000 01111000 01111000 01111000 \\

			8-GPSK Symbol 7 & 11110000 11110000 11110000 11110000 11110000 11110000 11110000 11110000 11110000 11110000 \\

			\hline

		\end{tabular}

\end{table*}

\subsection{Device efficiency}

\label{met:efficiency}

Our RA is based on the modulation of the reflected signal from each cell due to the variation of $\sigma$ with voltage. Ref.\citenum{TamagnoneFallahiMosigEtAl2014} demonstrated that specific upper bounds to the efficiency of such two-state modulators exist, and depend uniquely on the conductivity $\sigma_\mathrm{ON}$ and $\sigma_\mathrm{OFF}$ of graphene in the two states. If the corresponding cell reflection coefficients in the two states are $\Gamma_\mathrm{ON}$ and $\Gamma_\mathrm{OFF}$ then this bound is given by the following inequality\cite{TamagnoneFallahiMosigEtAl2014}:

\begin{align}\label{seqn:bound1}
\gamma_\mathrm{mod}\triangleq \frac{|\Gamma_\mathrm{ON}-\Gamma_\mathrm{OFF}|^2}{(1-|\Gamma_\mathrm{ON}|^2)(1-|\Gamma_\mathrm{OFF}|^2)} \nonumber\\ \leq
\gamma_\mathrm{R}\triangleq \frac{|\sigma_\mathrm{ON}-\sigma_\mathrm{OFF}|^2}{4\;\mathrm{Re}(\sigma_\mathrm{ON})\;\mathrm{Re}(\sigma_\mathrm{OFF})}
\end{align}

$\sigma$ at THz frequencies can be estimated analytically with the Drude model, as \cite{TamagnoneFallahiMosigEtAl2014}:

\begin{equation}\label{seqn:Drude}
\sigma_\mathrm{ON,OFF}=(R_\mathrm{ON,OFF}\;(1+j\omega\tau))^{-1}
\end{equation}

In our samples we have $R_\mathrm{ON}=800\;\Omega$, $R_\mathrm{OFF}=4000\;\Omega$, $\tau=45$ fs. From Eq. \ref{seqn:bound1} we get:

\begin{equation}\label{seqn:bounddrude}
\gamma_\mathrm{R} = \frac{(R_\mathrm{ON}-R_\mathrm{OFF})^2(1+\omega^2\tau^2)}{4\;R_\mathrm{ON}\;R_\mathrm{OFF}}\simeq 0.856
\end{equation}

For the RA, ideally, the reflection coefficients must have opposite phases and same absolute values. In practice this happens only approximately, and the actual deflected signal is proportional to the difference $D\triangleq\Gamma_\mathrm{ON}-\Gamma_\mathrm{OFF}$, while the sum $S\triangleq\Gamma_\mathrm{ON}+\Gamma_\mathrm{OFF}$ is responsible for an unwanted specular reflection component. Expressing the bound in $S$ and $D$ we get from Eq. \ref{seqn:bound1}:

\begin{align}\label{seqn:boundSD}
\gamma_\mathrm{mod}\;=\;\frac{16|D|^2}{(4-|S|^2-|D|^2)^2-4(\mathrm{Re}(SD^*))^2}\;\nonumber\\ \ge\;\frac{16|D|^2}{(4-|D|^2)^2}
\end{align}

and:

\begin{equation}\label{seqn:boundD}
\frac{16|D|^2}{(4-|D|^2)^2}\;\le\gamma_\mathrm{R}
\end{equation}

Eq. \ref{seqn:boundD} can now be inverted to find the maximum achievable $|D|$ with our SLG parameters. We get:

\begin{equation}\label{seqn:boundinv}
D\;\le\;|D|_\mathrm{max}\triangleq\sqrt{\frac{4(\gamma_\mathrm{R}+2-2\sqrt{\gamma_\mathrm{R}+1})}{\gamma_\mathrm{R}}}\;\simeq\;0.7832
\end{equation}

From the measurements in Fig.\ref{fig:Figure1}, $|D|\simeq 0.5$ at peak efficiency, below the computed $|D|_\mathrm{max}$. This is likely due to losses in the metal, therefore there is room for improvement in the cell design. The final power efficiency of the deflected beam is given by the average differential reflection coefficient ($|D|/2\simeq0.25$, i.e. $\sim-$12 dB, while the efficiency of the optimal device would be $\sim-8$ dB. $|D|_\mathrm{max}$ is a strict bound that cannot be exceeded\cite{TamagnoneFallahiMosigEtAl2014}, however its value can be increased using SLG having higher mobility or with gate oxide with better breakdown voltage and hence larger SLG tunability. The efficiency can be further increased by reducing the temperature to increase mobility and reduce the effect of thermal carriers.

\end{document}